\documentclass[aps,prx,twocolumn,showpacs,superscriptaddress,floatfix]{revtex4-1}  
\usepackage{ucs}
\usepackage{natbib}
\usepackage{graphicx}  
\usepackage{dcolumn}   
\usepackage{bbold}
\usepackage{bm}        
\usepackage{amssymb}   
\usepackage{amsmath}
\usepackage{color}
\usepackage{times}
\usepackage{verbatim}
\usepackage{mathrsfs}
\usepackage{hyperref}
\usepackage{ulem}
\hypersetup{colorlinks = True, urlcolor=blue, linkcolor=blue, citecolor=blue}

\newcommand{\bg}{\begin{pmatrix}}
\newcommand{\ed}{\end{pmatrix}}
\newcommand{\dg}{\dagger}

\newcommand{\kp}{\kappa}

\newcommand{\etal}{\textit{et al}.}

\usepackage{cancel}
\usepackage{comment}

\begin{document}

\title{Topological transitions in the  Yao-Lee spin-orbital model and effects of site disorder}
\author{Vladislav Poliakov}
\email{These two authors contributed equally to this work}
\affiliation{Department of Physics, Massachusetts Institute of Technology, Cambridge, MA 02139, USA}
\author{Wen-Han Kao} 
\email{These two authors contributed equally to this work}
\affiliation{School of Physics and Astronomy, University of Minnesota, Minneapolis, MN 55455, USA}
\author{Natalia B. Perkins} 
\email{nperkins@umn.edu}
\affiliation{School of Physics and Astronomy, University of Minnesota, Minneapolis, MN 55455, USA}

\begin{abstract}
The Yao-Lee model is an example of exactly solvable spin-orbital models that are generalizations of the original Kitaev honeycomb model with extra local orbital degrees of freedom. Similar to the Kitaev model, both spin and orbital degrees of freedom are effectively represented using sets of three-flavored Majorana fermions. The Yao-Lee model exhibits a quantum spin liquid ground state with gapped and immobile $Z_2$ fluxes and three-fold degenerate itinerant Majorana fermions. Our work demonstrated that by introducing different time-reversal symmetry (TRS) breaking fields one can split the degeneracy of Majorana fermions and close the gap for some of the bands, thus changing its topology. We calculated a comprehensive topological phase diagram for the Yao-Lee model by considering various combinations of TRS-breaking fields. This investigation revealed the emergence of distinct topological regions, each separated by nodal lines, signifying an evolution in the model's topological properties. We also investigated the impact of vacancies in the system. Our findings revealed that while vacancies modify the low-energy spectrum of the model, their presence has a limited impact on the topological properties of the model, at least for small enough concentrations.
\end{abstract}

\maketitle

\section{Introduction}
Over the past  decade, topology, frustration, and disorder
have been recognized as key ingredients to generate unconventional phases in solid-state systems with strong spin-orbit coupling (SOC) \cite{Balents2014}.
In a weakly correlated regime, it is now recognized that SOC plays a crucial role in the realization of topologically non-trivial phases of matter among
  which topological insulators are one of the most prominent examples \cite{KaneMele2005,Bernevig2006,Moessnerbook}. In a strongly correlated regime, 
  particularly in the presence of strong SOC,  quantum spin liquids (QSLs) 
   with exotic properties
such as fractionalization and long-range entanglement emerge as exceptionally captivating phases \cite{anderson1973resonating, 
Balents2010,Savary2016,Norman2016,KnolleMoessner2019,Broholm2020,Takagi2019}.
These properties of QSLs make them paradigmatic examples of phases endowed with topological order
\cite{Wen2002}.

 One of the  most famous examples of QSLs is realized  in  the
   Kitaev model on the honeycomb lattice \cite{Kitaev2006}. 
 This model  is of particular importance as it was the first
exactly-solvable model exhibiting a QSL ground state in two dimensions \cite{Kitaev2006}. Fractionalization of spin-1/2 degrees of freedom  into  $Z_2$ fluxes and Majorana fermions in the Kitaev model is strongly related to its underlying topological properties.  The Kitaev model complemented by
 the three-spin interactions arising from an external magnetic field realize topological phases with Majorana Chern numbers $C=0$ and $C=\pm 1$,
  which would be
distinguishable by their specific quantized values of the thermal Hall conductivity \cite{Kitaev2006}.

Remarkably, the bond-dependent Ising-like interactions, dubbed Kitaev interactions, 
naturally occur in 4d and 5d transition metal compounds with strong SOC and appropriate tri-coordinated geometry~\cite{Jackeli2009,Chaloupka2010},  providing a natural playground to study properties of the Kitaev QSL \cite{Kitaev2006}.
However, direct experimental observation and characterization of the Kitaev  QSL, and QSLs in general, is challenging since
the absence of the magnetic long-range order can happen for various reasons. For example, it can be caused by the presence of disorder, and since there are no clean samples in nature, it is crucial to understand whether the suppression of spin ordering is due to intrinsic properties of the QSL or generated by disorder within the material. 
It is also important to understand whether or not  the topological nature of the QSL survives up to a critical  strength of disorder or is being destroyed, for example,  by the formation of  topologically trivial random-singlet
phase \cite{Bhatt1982,Uematsu2018,Kimchi2018,Sanyal2021}.  
In general,
 QSLs are intrinsically less sensitive to disorder compared to topological insulators, as their topological order is not based on electronic band structures. Nevertheless,
  the impact of disorder should not be underestimated as it can still
    have non-trivial effects, influencing the nature of emergent excitations and modifying the characteristics of the QSL state.

Much of the intuition on the effect of disorder on the low-energy properties of  Kitaev QSL was obtained by using analogies with 
the disorder effects on the single-particle electron wavefunctions in solids \cite{Anderson1958,Lifshitz1965,Knolle2019,Nasu2020,Kao2021vacancy,Kao2021localization,Kao2022SDRG,Dantas2022}.
This is because the Kitaev model remains exactly solvable 
even in the presence of quenched disorder, thus being amenable for straightforward numerical analysis.
For example, it was shown that a small concentration of vacancies preserves most of the spin-liquid behavior in the Kitaev model but leads to distinct changes
 in the low-energy physics \cite{Knolle2019,Kao2021vacancy,Kao2021localization}. 
 In particular, vacancy-induced Majorana modes are accumulated
in a low-energy peak of the density of states whose form
across a broad window at low energies is consistent with a
power-law divergent density of states { \cite{Kao2021vacancy}}. Additionally, the states in this pileup are more localized than most of the other states in
the system \cite{Kao2021localization}. The localization of low-energy states is
particularly strong when time-reversal symmetry (TRS) is broken by the three-spin interactions arising from an external magnetic field. It was also shown that the Kitaev chiral QSL phase
survives up to a critical  concentration of  vacancies
due to a nontrivial flux
configuration, before being replaced by a topologically
trivial phase with power-law singularities \cite{Dantas2022}.

Nowadays, other models are known to be both exactly solvable and to possess QSL ground states  { \cite{YaoKivelson2009,Wang2009,Wu2009,Nakai2012,YaoLee2011,Carvalho2018,Seifert2020, Seifert2021,Natori2020,Janssen2021,Nica2023,Onur2023,Polyakov2023}}.
In this paper, we study the topology and disorder in the Yao-Lee  (YL) model \cite{YaoLee2011}. 
This model belongs to a  family of exactly solvable models that include additional local degrees of freedom, 
  such as orbital degrees of freedom \cite{YaoKivelson2009,Wang2009,Wu2009,YaoLee2011}, but still has a QSL ground state. {More recently, it has been shown that QSL ground states persist even in the higher-spin models\cite{Ma2023,Wu2024}.} 
  Similarly to the Kitaev model, spin and orbital degrees of freedom in this model can be represented in terms of Majorana fermions \cite{YaoLee2011}. However, differently from the Kitaev model, its fermionic representation includes three flavors of Majorana fermions. 
  The YL model exhibits an emergent $Z_2$ gauge symmetry with gapped $Z_2$ flux excitations defined exclusively in terms of the orbital degrees of freedom and fermionic excitations of the Majorana type related to spin
flips \cite{YaoLee2011,Carvalho2018,Seifert2020, Seifert2021,Natori2020,Janssen2021,Nica2023,Onur2023,Coleman2022,Tsvelik2022}.

 {The presence of spin and orbital degrees of freedom leads to a rich physics landscape, as it allows for a wide range of realistic microscopic perturbations. These perturbations
include onsite Zeeman magnetic fields as well as further
bond-dependent Kitaev-, off-diagonal  $\Gamma$-type and Dzyaloshinskii-Moriya  exchange
interactions \cite{Seifert2020,Seifert2021, Onur2023}.
These additional interactions can significantly alter the behavior of the YL model and lead to new phases of matter and unusual phase transitions.
For example, in the presence of perturbative nearest-neighbor (NN) Heisenberg interaction between  spin degrees of freedom, this model can be driven to a magnetically ordered state through a quantum critical point 
in the Gross-Neveu-SO(3) universality class \cite{Seifert2020,Seifert2021}. 
  The presence of orbital degrees of freedom in the YL model also leads to a bigger variety of topological phases compared with the original Kitaev model. 
  This is because there are multiple ways in which time-reversal symmetry-breaking terms can be constructed in the presence of orbital degrees of freedom \cite{Polyakov2023}, leading to a richer topological phase diagram compared to the original Kitaev model.}

   {In this work, we consider three different cases of time-reversal symmetry-breaking terms which keep the extended YL model exactly solvable:} the perturbative term does not leave the degeneracy of the three Majorana bands, the degeneracy is split to one lower-energy mode and two degenerate higher-energy modes or, vice versa, to two degenerate lower-energy modes and one higher energy mode. We show that depending
 on the nature of the TRS  breaking field, different topological phases are realized.

 The topological nature of the clean extended Yao-Lee model will also determine its response to  external disorder.  { Specifically, we examine the scenario where an 
  equal number of sites on both A and B sublattices become magnetically almost decoupled from the rest of the system. This is achieved by diminishing the exchange couplings between the defect site and its nearest neighbors. Colloquially, we will still call these sites 'vacancies' and the type of disorder 'site dilution'.}
We will show that similarly to the site-diluted Kitaev model \cite{Willans2010,Willans2011,Kao2021vacancy}, 
vacancies lead to the appearance of the zero-energy modes with quasilocalized wavefunctions on  the periphery of each  vacancy plaquette
 (on the other sublattice around the vacancy site), dubbed $p$-mode. In addition,  when the TRS-breaking field is relatively small, a single vacancy binds a flux. Thus, in the TRS-broken state, a vacancy pins topologically protected zero-energy  Majorana modes, dubbed $f$-modes, and the number of these topological modes depends on the nature of the TRS-breaking field. Recently, proposals have shown that these vacancy-induced modes can be potentially probed by scanning tunneling microscopy (STM) in various setups \cite{Kao2024short,Kao2024long,Takahashi2023}.

The rest of the paper is organized as follows: In Sec.~\ref{sec:model}, we introduce the Hamiltonian of the Yao-Lee model and the three types of TRS-breaking fields from the perturbation theory. In terms of the Majorana-fermion representation, the free-fermion band structure of the model and its dispersion relations are given analytically. In Sec.~\ref{sec:topology}, we provide the analytic formula to calculate the total Chern number of the model in the presence of TRS-breaking fields. We obtain the Chern-number phase diagram of the extended Yao-Lee model with different combinations of the TRS-breaking fields and all the topological transition lines. In Sec.~\ref{sec:vacancy}, we diagonalize the Majorana-fermion Hamiltonian in real space in order to study the effect of disorder. Specifically, we show how the presence of vacancies can modify the density of states and the Bott index, which is a topological index that can be computed in real space. While the low-energy density of states is significantly changed by the vacancy-induced eigenmodes, the Bott index behaves quite robustly against site dilution.
Finally, we present the concluding remarks in the last section.

\section{The model}\label{sec:model}

\begin{figure}
\center{\includegraphics[width=0.95\linewidth]{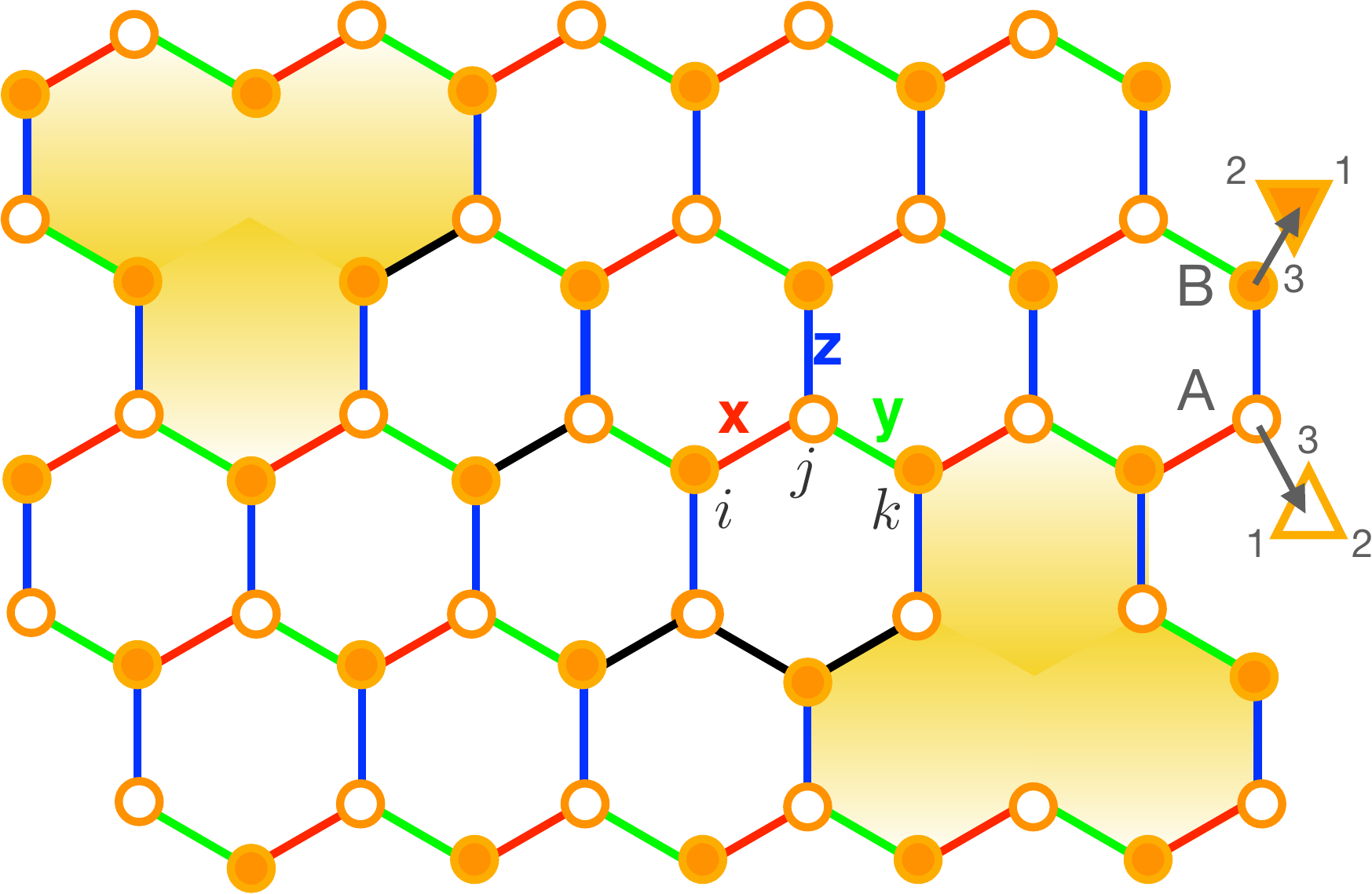}
\caption{The schematic representation of the  Yao-Lee model on the decorated honeycomb lattice \cite{YaoLee2011} with three types
of nearest-neighbor (NN) Ising bonds x,y, and z, shown by red, green, and blue lines. The triangles are labeled by $i$, $j$, and the sites within each triangle are labeled by 1, 2, and 3. A pair of vacancies is introduced on different sublattices A and B.  By flipping a string of link variables from $u=+1$
to $u=-1$ (shown by black bonds), a pair of fluxes can be attached to the pair of vacancies.}\label{YLlattice}}
\end{figure}

The  Yao-Lee model  \cite{YaoLee2011} has been proposed on the honeycomb lattice decorated with triangles (see Fig.\ref{YLlattice}):
\begin{equation}\label{original model}
    H_{\text{YL}}=\tilde{J}\sum_i \boldsymbol{S}_i^2+\sum_{<ij>_\lambda}J^\lambda_{ij} [\tau_i^\lambda \tau_j^\lambda][\boldsymbol{S}_i \boldsymbol{S}_j],
\end{equation}
where indices $i,j$ label the triangles (or the sites of the underlying honeycomb lattice), $ \boldsymbol{S}_i= \boldsymbol{S}_{i,1} +\boldsymbol{S}_{i,2}+ \boldsymbol{S}_{i,3}$ is  the total spin on each triangle, and the operators $\tau_i^\lambda$ describing the orbital degrees of freedom 
are defined as follows:  $\tau_i^x= 2(\boldsymbol{S}_{i,1}\boldsymbol{S}_{i,2}+1/4)$, $\tau_i^y= 2(\boldsymbol{S}_{i,1}\boldsymbol{S}_{i,3}-\boldsymbol{S}_{i,2}\boldsymbol{S}_{i,3})/\sqrt{3}$,  and  
$\tau_i^z= 4\boldsymbol{S}_{i,1}\cdot(\boldsymbol{S}_{i,2}\times\boldsymbol{S}_{i,3})/\sqrt{3}$.
If $\tilde{J}\gg J^\lambda$, where we assume that all the bonds of the same kind $\lambda$ have the same interaction strength $J^\lambda$, one can consider only the states with the total spin on each triangle being equal to  $\boldsymbol{S}=\frac{1}{2}\boldsymbol{\sigma}$. This also implies that $\tau_i^\lambda$ are Pauli matrices. With these constraints the first term in \eqref{original model} is a constant, and the second term  can be rewritten as
\begin{equation}\label{eff_ham}
    H_{\text{YL}}' =\frac{1}{4}\sum_{<ij>_\lambda}J^\lambda_{ij} [\tau_i^\lambda \tau_j^\lambda][\boldsymbol{\sigma}_i \boldsymbol{\sigma}_j].
\end{equation}
 Similarly to the Kitaev model \cite{Kitaev2006}, this model can be solved by the parton construction of the original degrees of freedom into Majorana fermions in an extended Hilbert space. In the Yao-Lee model, the spin and orbital operators are represented, respectively, as \cite{YaoLee2011}
\begin{equation}
\boldsymbol{\sigma}_i=-i{\bf c}_i\times
\boldsymbol{c}_{i},\,
\boldsymbol{\tau }_{i}=-{i}\boldsymbol{d}_{i}\times
\boldsymbol{d}_{i},
\end{equation}
where we use the vector notation for the Majorana fermion such that ${\bf c}_i= (c^{x}_i,c^{y}_i,c^{z}_i)$ and ${\bf d}_i= (d^{x}_i,d^{y}_i,d^{z}_i)$, along with the anticommutation relations 
$
    \{ c_i^\alpha,c_j^\beta\}=\{ d_i^\alpha,d_j^\beta\}=\delta_{ij} \delta_{\alpha,\beta}$ and $ \{ c_i^\alpha,d_j^\beta\}=0$. It is then  convenient to express Pauli matrices in terms of Majorana fermions:
\begin{equation}\label{PauliMajorana}
    \sigma_i^\alpha \tau_i^\beta=i c_i^\alpha d_i^\beta, \quad
    \sigma_i^\alpha=-\frac{\epsilon^{\alpha \beta \gamma}}{2}i c_i^\beta c_i^\gamma, \quad 
    \tau_i^\alpha=-\frac{\epsilon^{\alpha \beta \gamma}}{2}i d_i^\beta d_i^\gamma.
\end{equation}
 Substituting \eqref{PauliMajorana} into \eqref{eff_ham} gives the Hamiltonian in terms of Majorana fermions $\boldsymbol{c}_i$ and $Z_2$ gauge fields $\hat{u}_{ij}=-id_i^\lambda d_j^\lambda$:
\begin{equation}
    H_{\text{YL}}'
    =\frac{1}{4}\sum_{<ij>_\lambda}J^\lambda_{ij} \hat{u}_{ij} (i\boldsymbol{c_i} \boldsymbol{c_j}).
\end{equation}
 This Hamiltonian describes three flavors of free Majorana fermions ${c}_i^\kappa$ with $\kappa=x,y,z$, and they are coupled to the background of $Z_2$ gauge field $\hat{u}_{ij}$.  This  has a global SO(3) symmetry which rotates among the three species of Majorana
fermions, and it comes from the SU(2) symmetry of the original spin model \eqref{original model}.
Note that the Lieb theorem \cite{Lieb1994}  is applicable here as in the original Kitaev model \cite{Kitaev2006}, and it requires that the ground state lies in the zero-flux sector.
All three  Majorana fermions have the same dispersion in all flux sectors, i.e., the fermionic spectrum is three-fold degenerate in the whole Brillouin zone (BZ). In the ground-state zero-flux sector, the dispersion of each band is identical to the one obtained in the original Kitaev model \cite{Kitaev2006}, so that it is either gapless with the Dirac-like dispersion or gapped, depending on the ratio of the  coupling parameters
$J_{\lambda}$: $\varepsilon^\kappa_{ \mathbf{q}} = \sum_{\lambda =
x,y,z} J^{\lambda} e^{i \mathbf{q} \cdot \hat{\mathbf{r}}_{\lambda}}$
in terms of the three bond vectors $\hat{\mathbf{r}}_{x,y,z}$
pointing from any site in sublattice A to its respective neighbors
in sublattice B on the honeycomb lattice.

\subsection{ Time-reversal symmetry breaking}

{ We introduce time-reversal symmetry breaking in the Yao-Lee model by assuming
the pseudo-magnetic  "Zeeman-like" field acting on orbital degrees of freedom in combination with either Heisenberg or bond-anisotropic next-nearest-neighbor interactions between spin degrees of freedom. The idea that this pseudo-magnetic orbital field can be created by applying strain was first proposed theoretically for graphene \cite{Guinea2010} and later adopted for the Kitaev model \cite{Rachel2016}.

  To simplify our analysis, we choose the pseudo-magnetic Zeeman field  ${\mathbf{h}}$ in the [111] direction (triaxial strain). While the specific direction of the field is not crucial since it couples to all components of $\tau$ matrices, this choice ensures that all other lattice symmetries remain intact. The introduction of a Zeeman field in the Yao-Lee model breaks its exact solvability. However, as long as the fluxes are gapped and the strength of the field $h$ is sufficiently small and unable to excite them, we can treat this Zeeman term perturbatively. }

 There are several ways to introduce a TRS-breaking term that mimics the effect of a magnetic field but keeps the exact solvability of the model. 
 Following the original idea of Kitaev \cite{Kitaev2006},
 we perform the perturbation theory expansion relative to the zero-flux sector { (see  Appendix \ref{pert} for details of the derivation)}. This yields an effective Hamiltonian which preserves the gauge sector and therefore can still be represented in terms of the Majorana fermions $\boldsymbol{c}_i$ and the link variables $u_{ij}$.

Let us first assume that the Kitaev couplings are isotropic in strength, $J^x=J^y=J^z\equiv J$, and consider the perturbation in the form:
{
\begin{equation}\label{pertV}
     V=\sum_i (h_x \tau^x_i +h_y \tau^y_i +h_z \tau^z_i) + K \sum_{\langle\langle ik\rangle\rangle} \boldsymbol{\sigma}_i \boldsymbol{\sigma}_k,
\end{equation}
}
{ One can easily show (see  Appendix \ref{pert}) that the lowest-order term in the perturbation theory which breaks the time-reversal symmetry but does not change the flux sector and keeps Majorana fermion Hamiltonian quadratic in $c$ operators is the fourth-order term, which contains three $h$ and one $K$ terms}:
\begin{equation}\label{diagonal perturbation}
         H^{(4)}=12\sum_{ijk} \frac{K h_x h_y h_z}{\Delta_{jk}\Delta_{jk} \Delta_k} (\boldsymbol{\sigma}_i \boldsymbol{\sigma}_k )  \tau_i^x \tau_{j}^z \tau_k^y,
\end{equation}
where $\Delta_k$ is the change of energy after applying $\tau_k^y$ operator and $\Delta_{jk}$ is the same quantity but after applying $\tau_j^z \tau_k^y$. 


Another more general form of the  perturbation (\ref{pertV}) is 
{
\begin{equation}\label{pertValphabeta}
     V=\sum_i (h_x \tau^x_i +h_y \tau^y_i +h_z \tau^z_i) +  \sum_{\langle\langle ik\rangle\rangle}\sum_{\lambda,\mu} K_{\lambda\mu} \sigma_i^\lambda \sigma_k^\mu,
\end{equation}
}
where the matrix $K_{\lambda\mu}$ is given by 
\begin{equation}\label{Klambdamu}
K_{\lambda\mu}=K\delta_{\lambda\mu}+(1-\delta_{\lambda\mu})K'.
\end{equation}
In addition, we  will also consider the effect of 
a bond-dependent perturbation  
\begin{equation} \label{Klambdamunu}
K_{\lambda\mu}^{\nu}=K\delta_{\lambda\mu}+(1-\delta_{\lambda\mu})K'+(1-\delta_{\nu\mu})(1-\delta_{\nu\lambda})(1-\delta_{\lambda\mu}) K'',
\end{equation}
where index $\nu$ denotes the type of the second nearest-neighbor bond with
$\nu=x\,(y,z)$ when this bond is connected by the NN bonds of type $\mu=y\,(z,x)$ and $z\,(x,y)$, respectively (see Fig. \ref{YLlattice} for the illustration). 

{The effective Hamiltonian in the most general form is given by Eq.(\ref{h4}) in the Appendix
\ref{pert}.}
Using  the Majorana representation of the spins $\sigma$ and pseudospins $\tau$, 
  it can be written as
\begin{widetext}
\begin{eqnarray}\label{eq: Hamiltonian}
H_{\rm{eff}}=\sum_{\langle ij \rangle}\sum_{\lambda}i J_{ij}^\lambda \hat{u}_{ij}c_i^\lambda c_j^\lambda+
    \sum_{\langle\langle ik\rangle\rangle}\sum_{\lambda}i\kappa\; \hat{u}_{ij}\hat{u}_{jk} c_i^\lambda c_k^\lambda+
    \sum_{\langle\langle ik\rangle\rangle} \sum_{\lambda \neq \mu}i\eta\; \hat{u}_{ij}\hat{u}_{jk} 
    c_i^\lambda c_k^\mu
    +\sum_{\langle\langle ik\rangle\rangle \in {\nu}}
    \sum_{{ \nu}}
    i\chi\; \hat{u}_{ij}\hat{u}_{jk} 
            (c_i^\lambda c_k^\mu+c_i^\mu c_k^\lambda),
            \end{eqnarray}
\end{widetext}
where $\langle i j \rangle$ and $\langle \langle ik \rangle \rangle$ denote NN and  second NN bonds, respectively. In the last term, we call the second  NN bond  $\langle \langle ik \rangle \rangle$ of type
$\nu=x\,(y,z)$ when this bond is connected by the NN bonds $\langle i j \rangle$ and $\langle j k \rangle$ of type $\mu=y\,(z,x)$ and $z\,(x,y)$, respectively. Assuming that the flux gap is proportional to the isotropic coupling $J_{ij}^\lambda \equiv J$, as in the Kitaev model \cite{Kitaev2006}, the magnitude of the TRS-breaking interactions can be estimated as $\kappa\sim 6\frac{K h_x h_y h_z}{J^3}$, $\eta\sim 6\frac{K' h_x h_y h_z}{J^3}$ and $\chi\sim 6\frac{K'' h_x h_y h_z}{J^3}$.

\begin{figure*}
\includegraphics[width=0.45\textwidth]{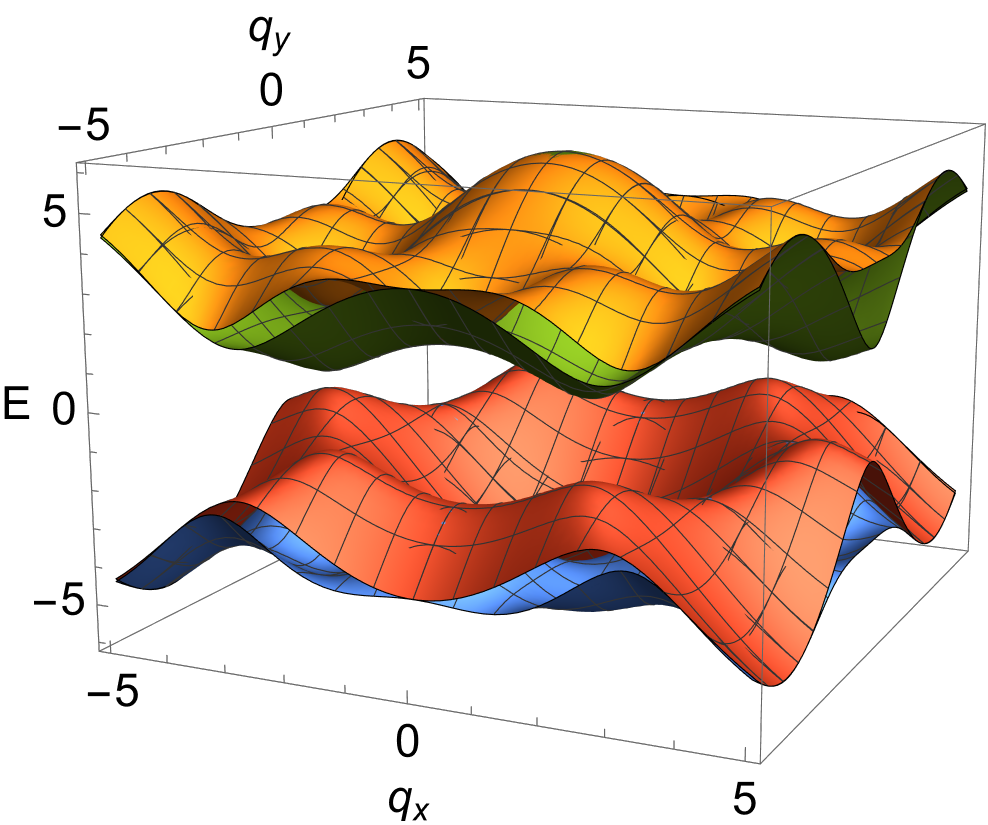}
\includegraphics[width=0.45\textwidth]{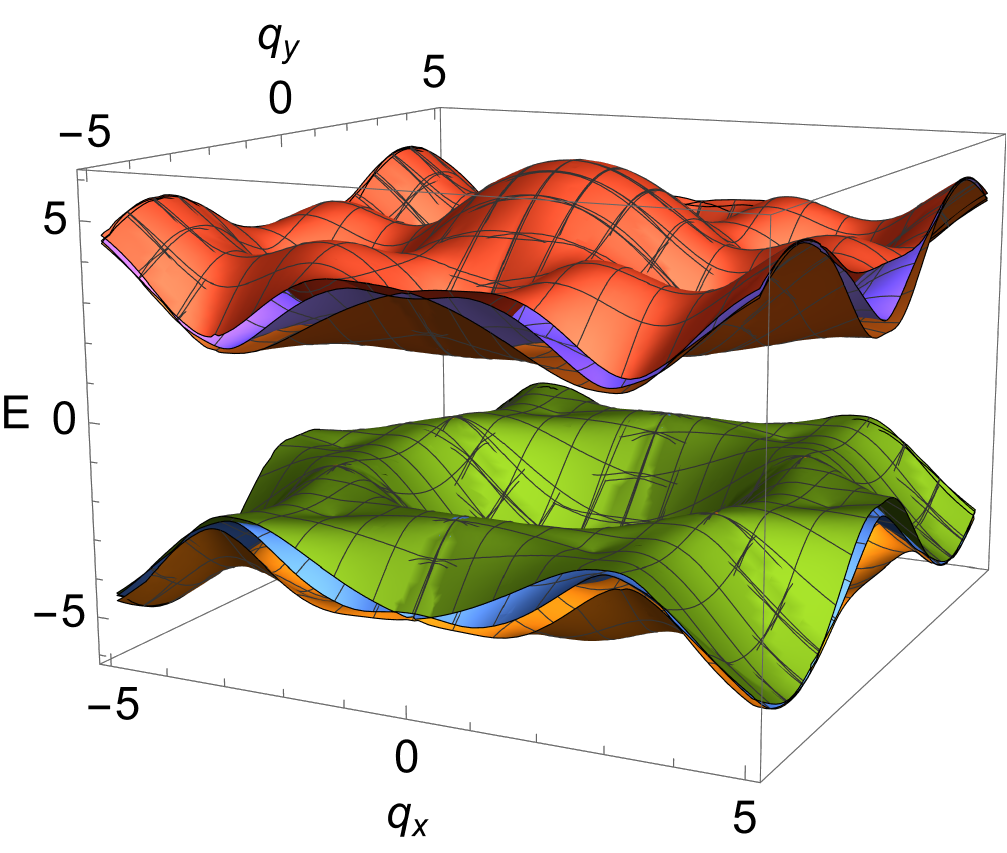}
\caption{\label{fig:spectrum}  The dispersion of the Majorana fermion's branches in the zero-flux ground state sector of the Yao-Lee model with broken TRS with (a) $\kappa=0.2$, $\eta=0.2$, $\chi=0.0$ and (b) $\kappa=0.3$, $\eta=0.0$, $\chi=0.4$. All energy scales
are given in the units of $J$ and all the second-neighbor hopping from the vacancy site are set to zero.}
\end{figure*}

\subsection{Majorana spectrum for $\kappa$- and $\eta$-interactions}
Let us first consider the case when $\chi=0$.
Within a given flux sector,  the $\hat{u}_{ij}$ operators are replaced by the corresponding eigenvalues ${u}_{ij}$, so the Hamiltonian (\ref{eq: Hamiltonian}) becomes quadratic in terms of the Majorana-fermion operators. 
Exploiting the bipartite nature of the honeycomb lattice, and noting
that each unit cell $l$ has two sites $\mathbf{r}^A_l$
 and
$\mathbf{r}^B_l$ in the two sublattices $A$ and $B$, the Hamiltonian (\ref{eq: Hamiltonian}) can be written as 
\begin{eqnarray}\label{HamMmatrix}
H_{\rm{eff}} =&&
\sum_{\langle l,l' \rangle }\sum_{\lambda }i M^\lambda_{ll'} c^\lambda_{A,l} c^\lambda_{B,l'} +\nonumber\\&&
\sum_{\langle\langle l,l' \rangle\rangle} \sum_{\lambda,\mu}i {\tilde M}_{ll'}^{\lambda\mu} (c^\lambda_{A,l} c^\mu_{A,l'} +c^\lambda_{B,l} c^\mu_{B,l'}),
\end{eqnarray}
 where  
 the first term describes the nearest-neighbor hopping of the Majorana fermions with 
 \begin{eqnarray}\nonumber
 M_{ll^\prime}^\lambda = -J_{ij}^\lambda
\hat{u}^{AB}_{ll'},
\end{eqnarray}
if the sites  $\mathbf{r}^A_l$ and
$\mathbf{r}^B_l$ are connected by  the $\lambda$-bond (we change the notations  for the bond variable from $\hat{u}_{ij}$ to $\hat{u}^{AB}_{ll'}$), and $M_{ll'}= 0$ otherwise; the second term describes the second-neighbor hopping
between two sites of the same sublattice with  
\begin{eqnarray}\nonumber
{\tilde M}_{ll^{\prime}}^{\lambda\mu}  =-\left[\kappa \delta_{\lambda\mu}+\eta(1-\delta_{\lambda\mu})\right] \hat{u}^{AB}_{ll^{\prime\prime}} \hat{u}^{AB}_{l^{\prime} l^{\prime\prime}}.
\end{eqnarray}

 The free-fermion Hamiltonian (\ref{HamMmatrix}) can be diagonalized in the momentum space if the system remains translational invariant. {Even in the presence of the second-neighbor hopping, the zero-flux sector is still the ground state.
Hence, we  can set all link variables $u^{AB}=1$, so that the Hamiltonian in momentum space reads}
\begin{eqnarray}  
H_{\rm{eff}}=&\frac{1}{2}\sum_{\bf q} {\bf c}^T_{-\boldsymbol{q}} \,  i{\mathcal H}_{\boldsymbol{q}}\,{\bf c}_{\boldsymbol{q}},
\end{eqnarray}
where ${\bf c}^T_{\boldsymbol{q}}=\left(c^x_{A,\boldsymbol{q}},\,c^y_{A,\boldsymbol{q}},\,c^z_{A,\boldsymbol{q}},\,c^x_{B,\boldsymbol{q}},\,c^y_{B,\boldsymbol{q}},\,c^z_{B,\boldsymbol{q}}\right)
$ and
the structure of the  matrix  $i{\mathcal H}_{\boldsymbol{q}}$ is given by
 \begin{equation}
       i{\mathcal H}_{\boldsymbol{q}}=
        \begin{pmatrix}
            M_{\boldsymbol{q}} & F_{\boldsymbol{q}} & F_{\boldsymbol{q}}\\
            F_{\boldsymbol{q} }& M_{\boldsymbol{q} }& F_{\boldsymbol{q}}\\
            F_{\boldsymbol{q}} & F_{\boldsymbol{q}} & M_{\boldsymbol{q}}
        \end{pmatrix}
    \end{equation}
  with
    \begin{eqnarray}\nonumber
        &&M_{\boldsymbol{q}}=
        \begin{pmatrix}
            \Delta_{\kappa}(\boldsymbol{q}) & i f(\boldsymbol{q})\\
            -i f^*(\boldsymbol{q}) & -\Delta_{\kappa}(\boldsymbol{q})
        \end{pmatrix}, \\
       && F_{\boldsymbol{q}}=
        \begin{pmatrix}
            \Delta_{\eta}(\boldsymbol{q})/2 & 0\\
            0 & -\Delta_{\eta}(\boldsymbol{q})/2
        \end{pmatrix}.
    \end{eqnarray}
 In the isotropic limit,  $f(\boldsymbol{q})=2J (1+e^{i\boldsymbol{q}\cdot \boldsymbol{n}_1}+e^{i\boldsymbol{q}\cdot\boldsymbol{n}_2})$, where 
${\bf n}_1=(\frac{1}{2},\frac{\sqrt{3}}{2})$ and ${\bf n}_2=(-\frac{1}{2},\frac{\sqrt{3}}{2})$  are two unit vectors of the Bravais lattice of the honeycomb lattice.
Other terms describe the second-neighbor Majorana hopping due to the time-reversal symmetry-breaking fields: $\Delta_{\kappa}(\boldsymbol{q})=4\kappa[\sin(\boldsymbol{q}\cdot\boldsymbol{n}_1)-\sin(\boldsymbol{q}\cdot\boldsymbol{n}_2)+\sin(\boldsymbol{q}\cdot(\boldsymbol{n}_2-\boldsymbol{n}_1)]$, $\Delta_{\eta}(\boldsymbol{q})=4\eta[\sin(\boldsymbol{q}\cdot\boldsymbol{n}_1)-\sin(\boldsymbol{q}\cdot\boldsymbol{n}_2)+\sin(\boldsymbol{q}\cdot(\boldsymbol{n}_2-\boldsymbol{n}_1)]$.

When  $\kappa=\eta=0$ and the time-reversal symmetry is not broken,  the Majorana fermion spectrum $\varepsilon_{i,\boldsymbol{q}}=|f(\boldsymbol{q})|$ contains three degenerate branches, each with two Dirac points at the corners of the Brillouin zone $\pm \mathbf{K}$.
If  $\kappa\neq 0$ but $\eta=0$, then all three branches are gapped at the Dirac points but remain degenerate.
Thus,  when $\eta=0$, the spectrum of the Yao-Lee model is nothing else but three copies of the spectrum of the Kitaev model.
However, if $\eta\neq 0$, the Majorana modes hybridize with each other, and the degeneracy is partially lifted (two of them remain degenerate). The eigenvalues of the perturbed Yao-Lee model (\ref{eq: Hamiltonian}) are given by
\begin{align}\label{spectrum}
\begin{split}
    &\varepsilon_1 (\boldsymbol{q})=\pm\sqrt{(\kappa+\eta)^2\Delta^2(\boldsymbol{q})+|f(\boldsymbol{q})|^2}\\
    &\varepsilon_2 (\boldsymbol{q})= \pm\sqrt{(\kappa-\frac{\eta}{2})^2\Delta^2(\boldsymbol{q})+|f(\boldsymbol{q})|^2}\\
    &\varepsilon_3 (\boldsymbol{q})= \pm\sqrt{(\kappa-\frac{\eta}{2})^2\Delta^2(\boldsymbol{q})+|f(\boldsymbol{q})|^2},
\end{split}
\end{align}
where $\Delta(\boldsymbol{q})=4[\sin(\boldsymbol{q}\cdot\boldsymbol{n}_1)-\sin(\boldsymbol{q}\cdot\boldsymbol{n}_2)+\sin(\boldsymbol{q}\cdot(\boldsymbol{n}_2-\boldsymbol{n}_1)]$.

From the above expressions, we can see that while the first branch is gapped for all non-zero $\kappa$ and $\eta$, the other two branches can remain gapless even when { the time-reversal} symmetry is broken. { In Fig.\ref{fig:spectrum}(a), we show an example of the Majorana band structures for $\kappa=\eta=0.2$. With $\kappa$- and $\eta$-interactions, only two separated bands are visible in the band structure, because one of them is doubly degenerate.}
{Based on Eq.(\ref{spectrum})}, the Majorana gap  at the Dirac points $\pm \mathbf{K}$ is  determined by
    \begin{equation}
        \Delta_M =
        6\sqrt{3}|\kappa-\eta/2|,
    \end{equation}
so it vanishes when $\kappa=\eta/2$.

\subsection{Majorana spectrum for $\kappa$-, $\eta$-, and $\chi$-interactions}
When all three types of the TRS-breaking fields are present in the system, the Hamiltonian matrix  in the  momentum space reads
 \begin{equation}
       i{\mathcal H}_{\boldsymbol{q}}=
        \begin{pmatrix}
            M_{\boldsymbol{q}} & F_{\boldsymbol{q}}+Z_{\boldsymbol{q}} & F_{\boldsymbol{q}}+Y_{\boldsymbol{q}}\\
            F_{\boldsymbol{q} }+Z_{\boldsymbol{q}}& M_{\boldsymbol{q} }& F_{\boldsymbol{q}}+X_{\boldsymbol{q}}\\
            F_{\boldsymbol{q}}+Y_{\boldsymbol{q}} & F_{\boldsymbol{q}}+X_{\boldsymbol{q}} & M_{\boldsymbol{q}}
        \end{pmatrix}
    \end{equation}
  with additional terms
\begin{align}
\begin{split}
&X_{\boldsymbol{q}}=
        \begin{pmatrix}
            \Delta_\chi^{x}(\boldsymbol{q})/2 & 0\\
            0 & -\Delta_\chi^{x}(\boldsymbol{q})/2
        \end{pmatrix}\\
&Y_{\boldsymbol{q}}=
        \begin{pmatrix}
            \Delta_\chi^{y}(\boldsymbol{q})/2 & 0\\
            0 & -\Delta_\chi^{y}(\boldsymbol{q})/2
        \end{pmatrix}\\
&Z_{\boldsymbol{q}}=
        \begin{pmatrix}
            \Delta_\chi^{z}(\boldsymbol{q})/2 & 0\\
            0 & -\Delta_\chi^{z}(\boldsymbol{q})/2
        \end{pmatrix},
\end{split}
\end{align}
where we denote
\begin{align}
\begin{split}
&\Delta_\chi^{x}(\boldsymbol{q})=-4\chi \sin(\boldsymbol{q}\cdot\boldsymbol{n}_2)\\
&\Delta_\chi^{y}(\boldsymbol{q})=4\chi \sin(\boldsymbol{q}\cdot\boldsymbol{n}_1)\\
&\Delta_\chi^{z}(\boldsymbol{q})=4\chi \sin[\boldsymbol{q}\cdot(\boldsymbol{n}_2-\boldsymbol{n}_1)].
\end{split}
\end{align}

Diagonalizing the Hamiltonian $i{\mathcal H}_{\boldsymbol{q}}$,  we obtain the spectrum for Majorana fermions. While the explicit formula for the spectrum looks somewhat complex, we can still find the gap-closing conditions analytically. In the presence of  $\chi$ interaction, the spectrum shows a Mexican-hat behavior near the Dirac points as shown in Fig.\ref{fig:spectrum}(b). 
At these critical points, the parameters $\Delta^{x}_{\chi}(\boldsymbol{q})=\Delta^{y}_{\chi}(\boldsymbol{q})=\Delta^{z}_{\chi}(\boldsymbol{q})=2\sqrt{3}\chi$, leading to the closure of the gap in two branches of the spectrum when 
\begin{eqnarray}\label{first two branches}
\kappa-\eta/2-\chi/6=0.
\end{eqnarray}
The  gap in the third branch closes when
\begin{eqnarray}\label{third branch}
\kappa+\eta+\chi/3=0 .
\end{eqnarray}

\begin{figure*}
     \includegraphics[width=1.0\textwidth]{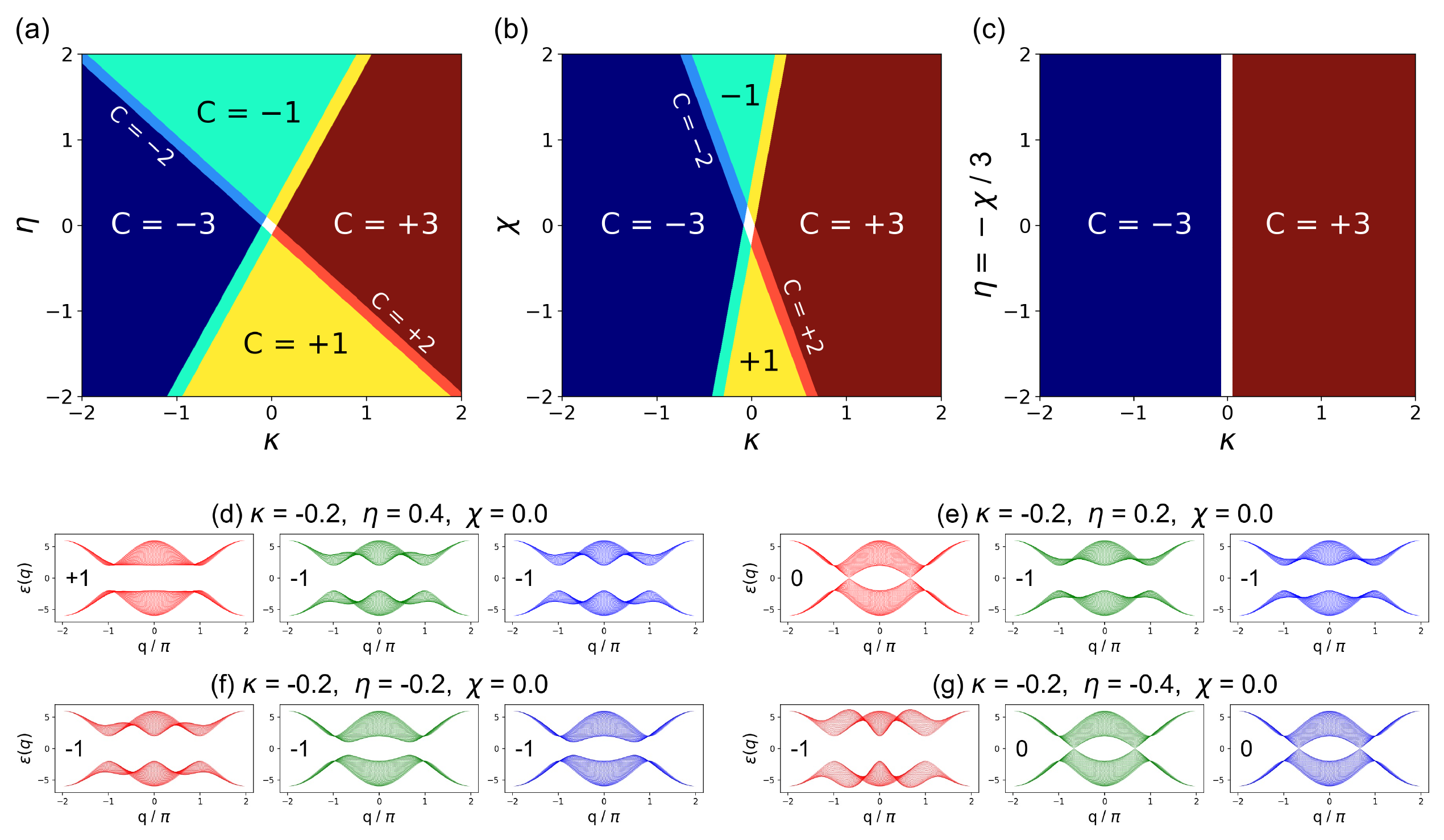}
                   \caption{\label{fig:phase diagram}  (a-c) Chern-number phase diagram of the extended Yao-Lee model, based on the analytical expressions $C = C_1 + C_2 + C_3$ where $C_1= \operatorname{sign} (\kappa+\eta+\frac{\chi}{3}) $ and  
        $C_2=C_3=\operatorname{sign} (\kappa -\frac{\eta}{2}-\frac{\chi}{6})$. The white dot in (a) and (b) and the white line in (c), represent the time-reversal symmetric case with $C = 0$. The colored oblique lines in (a) and (b) represent the critical lines. (d-g) The fermionic band structures for four points in the phase diagram (a). The number on each plot labels the Chern number of the occupied band. In (e), one of the three bands becomes gapless but the other two remain gapped. In (g), two bands become gapless and the other remain gapped. All energy scales
are given in the units of $J$.}
\end{figure*}

\section{Topological structure of the spectrum}\label{sec:topology}

 In this section, we explore the topological nature of the phases realized in the extended Yao-Lee model in the presence of the TRS-breaking interactions discussed in the previous section. 
 An important topological quantity characterizing a two-dimensional system of non-interacting fermions with an energy gap is the  Chern number. Since there are three Majorana bands in the Yao-Lee model,
 the total Chern number of the system is the sum of the Chern numbers of all negative-energy Majorana bands.

As the first step, we write down eigenvectors for the negative-energy eigenvalues in Eq.(\ref{spectrum}).  It is worth noting a key observation that for a given set of parameters ($\kappa$, $\eta$, $\chi$), it is possible to smoothly tune $\chi$ to zero while keeping $\left(\eta+\chi/3\right)$ constant. This process does not involve the gap closing and thus the total Chern number remains the same. Therefore, it is sufficient to calculate the Chern number solely for the parameters ($\kappa$, $\eta$) and subsequently make the substitution $\eta \to \eta+\chi/3$. 
For the Hamiltonian  with only $\kappa$ and $\eta$ TRS-breaking terms, it is convenient to introduce the following notations:
\begin{align}
\begin{split}
&g_{\gamma}(\boldsymbol{q})=\frac{-\gamma \Delta(\boldsymbol{q})+\sqrt{\gamma^2\Delta^2(\boldsymbol{q})+|f(\boldsymbol{q})|^2}}{f^*(\boldsymbol{q})}\\
&G_{\gamma}(\boldsymbol{q}) = \frac{g_{\gamma (\boldsymbol{q})}}{\sqrt{1+|g_{\gamma}(\boldsymbol{q})|^2}}\\
&F_{\gamma}(\boldsymbol{q}) = \frac{1}{\sqrt{1+|g_{\gamma}(\boldsymbol{q})|^2}},
\end{split}
\end{align}
where $\gamma = \gamma(\kappa,\eta)$ is a function of the two parameters. Then the energy eigenvectors can be expressed as
\begin{align}\label{eigenvectors}
\begin{split}
&w_{1,{\boldsymbol{q}}}=\frac{1}{\sqrt{3}}
        \begin{pmatrix}
            -i G_{\kappa+\eta}(\boldsymbol{q})\\
            F_{\kappa+\eta}(\boldsymbol{q})\\
            -i G_{\kappa+\eta}(\boldsymbol{q})\\
            F_{\kappa+\eta}(\boldsymbol{q})\\
            -i G_{\kappa+\eta}(\boldsymbol{q})\\
            F_{\kappa+\eta}(\boldsymbol{q})\\
        \end{pmatrix}\\
&w_{2,{\boldsymbol{q}}}=\frac{1}{\sqrt{2}}
        \begin{pmatrix}
            i G_{\kappa-\frac{\eta}{2}}(\boldsymbol{q})\\
            -F_{\kappa-\frac{\eta}{2}}(\boldsymbol{q})\\
            0\\
            0\\
            -i G_{\kappa-\frac{\eta}{2}}(\boldsymbol{q})\\
            F_{\kappa-\frac{\eta}{2}}(\boldsymbol{q})\\
        \end{pmatrix}\\
&w_{3,{\boldsymbol{q}}}=\frac{1}{\sqrt{2}}
        \begin{pmatrix}
            i G_{\kappa-\frac{\eta}{2}}(\boldsymbol{q})\\
            -F_{\kappa-\frac{\eta}{2}}(\boldsymbol{q})\\
            -i G_{\kappa-\frac{\eta}{2}}(\boldsymbol{q})\\
            F_{\kappa-\frac{\eta}{2}}(\boldsymbol{q})\\
            0\\
            0\\
        \end{pmatrix}.
\end{split}
\end{align}
The single-band Chern number is defined as the integral of Berry curvature over the Brillouin zone \cite{Thouless1982,Avron1983}:
    \begin{eqnarray}\label{cherndef}
        C_n=\frac{-i}{2\pi}\int_{\mathrm{BZ}} dq_x dq_y (\frac{\partial w_{n\alpha}^*}{\partial q_x}\frac{\partial w_{n\alpha}}{\partial q_y}-\frac{\partial w_{n\alpha}^*}{\partial q_y}\frac{\partial w_{n\alpha}}{\partial q_x}),
    \end{eqnarray}
where $w_{n\alpha}$, in the case of the Yao-Lee model, are the components of the six-dimensional vector $w_n$. By substituting eigenvectors (\ref{eigenvectors}) into  (\ref{cherndef}), we  find the Chern numbers for the $n$-th band:
    \begin{eqnarray}
        C_n=\frac{1}{\pi} \operatorname{Im} \int_{\mathrm{BZ}} dq_x dq_y \frac{\partial G^{*}_{\gamma_n}(\boldsymbol{q})}{\partial q_x} \frac{\partial G_{\gamma_n}(\boldsymbol{q})}{\partial q_y},
    \end{eqnarray}
    where $\gamma_1 = \kappa+\eta$ and $\gamma_2 = \gamma_3 = \kappa - \eta/2$. This integral is similar to the Chern number of the Kitaev honeycomb model under the similar notations \cite{Kitaev2006}:
    \begin{align}
    \begin{split}
           &i{\mathcal H}^{\text{Kitaev}}_{\boldsymbol{q}}=
        \begin{pmatrix}
            \Delta_{\kappa}(\boldsymbol{q}) & i f(\boldsymbol{q})\\
             -i f^*(\boldsymbol{q}) & -\Delta_{\kappa}(\boldsymbol{q})
        \end{pmatrix}\\
    &w^{\text{Kitaev}}_{\boldsymbol{q}}=\frac{1}{\sqrt{1+|g_{\kappa}(\boldsymbol{q})|^2}}
        \begin{pmatrix}
            -i g_{\kappa}(\boldsymbol{q})\\
            1\\
        \end{pmatrix}\\
        &C^{\text{Kitaev}}=\frac{1}{\pi} \operatorname{Im} \int_{\mathrm{BZ}} dq_x dq_y \frac{\partial G^{*}_{\kappa}(\boldsymbol{q})}{\partial q_x} \frac{\partial G_{\kappa}(\boldsymbol{q})}{\partial q_y},
    \end{split}
    \end{align}
    and thus $C^{\text{Kitaev}} = \mathrm{sign}\Delta_0$ where $\Delta_0\equiv\Delta_{\kappa}(\boldsymbol{\bf K})$ is the value of $\Delta_\kappa$ at one of the Dirac points, and $\Delta_{\kappa}(-\boldsymbol{\bf K})=-\Delta_{\kappa}(\boldsymbol{\bf K})$ at the other Dirac point $-\boldsymbol{\bf K}$. Since the expressions of the Berry curvature in the Kitaev honeycomb model and in the Yao-Lee model  
    only differ by the function $\gamma(\kappa,\eta)$,   the Chern numbers    for the bands  in the Yao-Lee model can be expressed as:
    \begin{align}
    \begin{split}
        &C_1= \operatorname{sign} (\kappa+\eta),\\
        &C_2=\operatorname{sign} (\kappa -\frac{\eta}{2}),\\
        &C_3=\operatorname{sign} (\kappa -\frac{\eta}{2}).
    \end{split}
    \end{align}
By making the substitution $\eta \to \eta+\chi/3$,
the Chern numbers  in  the Yao-Lee model with  $\kappa$, $\eta$ and $\chi$ TRS-breaking interactions are given by
    \begin{align} \label{C1C2C3}
    \begin{split}
        &C_1= \operatorname{sign} (\kappa+\eta+\frac{\chi}{3}),\\
        &C_2=\operatorname{sign} (\kappa -\frac{\eta}{2}-\frac{\chi}{6}),\\
        &C_3=\operatorname{sign} (\kappa -\frac{\eta}{2}-\frac{\chi}{6}).
    \end{split}
    \end{align}

In Fig.~\ref{fig:phase diagram}(a),  the topological phase diagram of the Yao-Lee model
 with  $\kappa$ and $\eta$  varying between $-$2 and 2 and $\chi=0$
is computed
based on the total Chern number $C = C_1+C_2+C_3$ from the analytical expressions  (\ref{C1C2C3}); the Chern number formula (\ref{C1C2C3}) and phase diagrams are further confirmed by the numerical calculations on a discretized Brillouin zone with the efficient method introduced by Fukui \etal ~\cite{Fukui2005}. 

{The phase diagram has four different topological phases, characterized by $ C=\pm 3$ and $\pm 1$, and four critical lines between the adjacent phases. At the center of the phase diagram with $\kappa=\eta=\chi=0$, the total Chern number simply vanishes due to the absence of any TRS-breaking fields.}
In the special case where $\eta=\chi = 0$, the spectrum of Majorana fermions simplifies to a three-fold replica of the spectrum in the Kitaev model, leading to a total Chern number of $C = 3\operatorname{sign}(\kappa)$.
 For finite values of $\eta$ or $\chi$, the Majorana bands of different flavors may have distinct Chern numbers.
 For example, in Fig.~\ref{fig:phase diagram}(d),  for   { $\kappa = -0.2$ and $\eta=0.4$}, the two degenerate bands (colored in green and blue) have $C_2=C_3 = -1$ and the other one (colored in red) has $C_1 = +1$. Therefore, the total Chern number is $C = -1$.   
 
 { Figs.~\ref{fig:phase diagram}(d)-(g)} depict the band structures of the three Majorana flavors for { four} specific points
  on the phase diagram in Fig.~\ref{fig:phase diagram}(a). Note that as long as $\chi=0$, two of the bands remain degenerate. Additionally, when some Majorana bands become gapless, signifying the emergence of Dirac points, the Chern number of the corresponding bands becomes zero. This behavior is visually represented in Figs.~\ref{fig:phase diagram}{ (e) and (g)}.  Consequently, the gap closing in the non-degenerate band or in the two-fold degenerate bands manifests, respectively, in a change in the total Chern number by 1 or by 2. This change subsequently gives rise to the emergence of { critical lines} delineating different topological phases in the phase diagram.

In Fig.\ref{fig:phase diagram}(b), we examine topological transitions in the presence of both $\kappa$ and $\chi$ interactions. 
  Since   $\eta$ and $\chi$  have similar effects on the band structure, the Chern-number phase diagram of Fig.~\ref{fig:phase diagram}(b)  resembles Fig.~\ref{fig:phase diagram}(a), but with modified phase boundaries.  Considering the simultaneous effect of both $\eta$ and $\chi$ interactions does not qualitatively change the picture, however,
 at the fine-tuned line  $\chi=-3\eta$, their effects are exactly canceled out. Consequently, the phase diagram mirrors that of three copies of the Kitaev honeycomb model. This distinct scenario is illustrated in Fig.~\ref{fig:phase diagram}(c){, where the critical line in the middle denotes $\kappa = 0$ and thus $C = 0$}.

\section{ Vacancies in the Yao-Lee model}\label{sec:vacancy}

Robustness against perturbations is a fundamental aspect of topological phenomena.  
  Hence, in this section, we will investigate how the topological phase diagram of the clean Yao-Lee model with broken TRS is modified in the presence of vacancies.
 Here we assume that a vacancy describes a magnetic moment very weakly connected to the rest of the system due to extremely strong but relatively rare bond randomness. While this is not the most general definition, it describes well the situation that might happen in the materials of interest.

{
To introduce vacancies into the isotropic  Yao-Lee model, one needs to randomly pick several lattice sites in Hamiltonian \eqref{original model} and reduce coupling to their neighbors: $J_{ij}\rightarrow J'$. Since obtaining Majorana fermion Hamiltonian is absolutely analogous to the clean case, we can immediately rewrite the first term in (\eqref{eq: Hamiltonian}) as }
\begin{equation}\label{eq:HamMFvac}
   i\sum_{\substack{\left \langle ij \right \rangle\\ i,j \in \mathbb{P}}}\sum_\lambda
  J\hat{u}_{ij }c^\lambda_{i}c^\lambda_{j} + i\sum_{\substack{\left \langle kl \right \rangle\\ k \in \mathbb{V}, l \in \mathbb{P}}}\sum_\lambda
   J^{\prime}\hat{u}_{ kl }c^\lambda_{k}c^\lambda_{l},
\end{equation}
where $\mathbb{P}$ denotes the subset of normal lattice sites and $\mathbb{V}$ denotes the subset of vacancy sites (see Fig.\ref{YLlattice}). {The rest of the terms in \eqref{eq: Hamiltonian} should be treated more carefully. In the remaining part of the paper, we set all the second-neighbor hopping from and across the vacancy site to zero (a detailed discussion can be found in Appendix \ref{flux binding})}.
\par We consider a 
compensated case with equal numbers of vacancies on the two sublattices of the honeycomb lattice. In the limit of $J^{\prime} \ll J$, the sites belonging to $\mathbb{V}$ behave as quasivacancies. 
As $J^{\prime}$ approaches zero, these sites transform into true vacancies. 
In this case, the Majorana fermions ${c}^\alpha$ remain on the vacancy site, but their nearest-neighbor hopping amplitudes are removed. 
In the presence of vacancies,   the Hamiltonian (\ref{eq: Hamiltonian}) can still be diagonalized 
 but only
 numerically \cite{Kao2021vacancy}. Performing diagonalization on finite-size clusters with periodic boundary conditions, we obtain  the diagonalized Hamiltonian in the form
 \begin{align}\label{eq:HamMF-diag}
\mathcal{H} = \sum_{n}\epsilon_{n}(\psi_{n}^{\dagger}\psi_{n}-\frac{1}{2}),
\end{align}
where the eigenmodes $\psi_n$ with fermionic energies $\epsilon_n \equiv\epsilon_n \left (\{ J_{ ij }\},\{ u_{ij}\}\right)$ for a given realization of disorder and a given flux sector
are {obtained through the Bogoliubov transformation in real space}
\begin{align}\label{XY}
 \bg \psi \\ \psi^{\dg} \ed = \bg X^* & Y^* \\ Y & X \ed \bg f \\ f^{\dg} \ed \equiv T^{\dg}\bg f \\ f^{\dg} \ed,
\end{align}
with complex fermions $f$ defined from { the combinations of} the original Majorana fermions $c_{A/B}= (c^{x}_{A/B},c^{y}_{A/B},c^{z}_{A/B})$:
\begin{align}
\bg f \\ f^{\dg} \ed = \frac{1}{2}\bg \mathbb{1} & i\mathbb{1} \\ \mathbb{1} & -i\mathbb{1}\ed \bg c_A \\ c_B \ed,
\end{align}
{ where $\mathbb{1}$ is the identity matrix with each dimension being equal to the number of unit cells.}

\subsection{Density of states, inverse participation ratio and localization of the low-energy states}  \label{DOS}

Next, we explore how the presence of vacancies affects
the low-energy states in the Yao-Lee model. First, we recall that in the Kitaev spin liquid, introducing vacancies leads to the emergence of nearly zero-energy eigenmodes with quasilocalized wavefunctions around each vacancy site \cite{Willans2010,Willans2011,Kao2021vacancy}. Once TRS is broken, the chiral Kitaev spin liquid exhibits additional low-energy modes associated with fluxes bound to vacancies. These localized modes exist in the bulk gap opened by the magnetic field, hybridize with each other, and become disconnected from the bulk modes under strong-enough fields \cite{Kao2021vacancy, Kao2021localization}. Here, we extend this exploration to the Yao-Lee model with vacancies.

 To characterize the low-energy states, we calculate the density of states (DOS) defined as $N(E) = \sum_{n}\delta(E-\epsilon_{n})$ and present the disorder-averaged results over 500 independent site-disordered samples. Compared to the clean model, only the DOS at the low energies will be drastically modified by the vacancy-induced modes. Furthermore, we use the inverse participation ratio (IPR), $\mathcal{P}_{n} = \sum_{i}|\phi_{n,i}|^{4}$, to quantify the strength of the localization for all eigenmodes. In the above definition, the index $n$ labels the eigenmode wavefunction $\phi_{n,i}$, and the index $i$ labels the lattice site. For a delocalized mode, the IPR scales roughly as $\sim 1/N$ in a system with $N$ sites, reflecting the uniform spreading of the wavefunction over the entire lattice. On the other hand, for a localized mode, the IPR remains finite as the wavefunction is confined to a small portion of the lattice.

\begin{figure}
\includegraphics[width=1.0\linewidth]{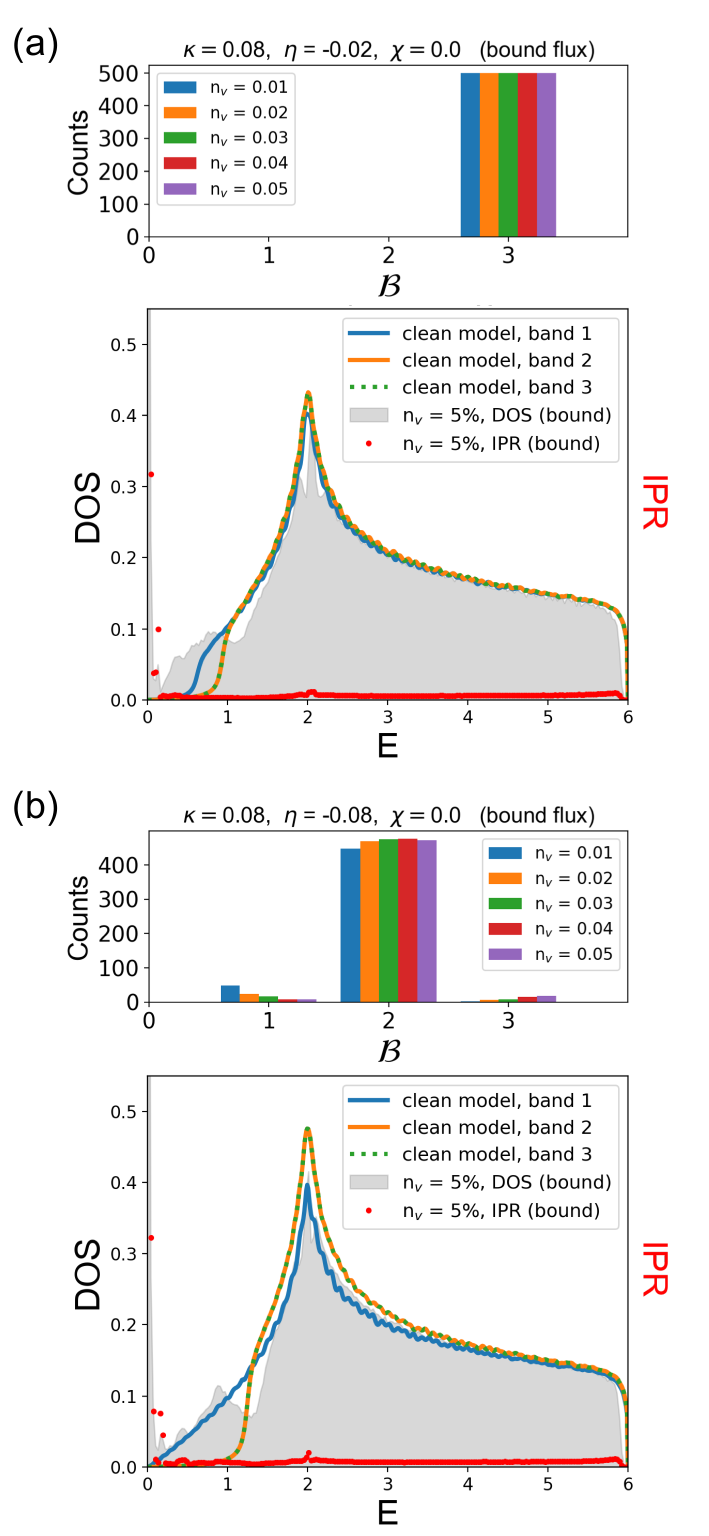}
\caption{\label{fig:DOS_vac} 
The {bar chart} of the Bott index, disordered averaged density of states (DOS), and disordered averaged inverse participation ratio (IPR). The disorder data was calculated from 500 random samples with L = 20 in the bound-flux sector. { In the bar chart, all the data points have integer-valued Bott index, and the value associated with each bar is shown in the nearest tick of the horizontal axis}. In the DOS plot, three curves representing the DOS of the three bands from the clean model are given as references. In (a), all three bands of the clean model are gapped and two of them are equivalent. In (b), one band becomes gapless since the parameters represent the phase-boundary value. All energy scales are given in the units of $J$ and all the second-neighbor hopping around the vacancies are set to zero { (see Appendix \ref{flux binding})}.}
\end{figure}

In Fig.~\ref{fig:DOS_vac}, we compare the DOS for the clean and the site-disordered Yao-Lee model with the vacancy concentration $n_v = 5\%$. For the $L = 20$ system, it contains $2L^2\times n_v = 40$ vacancies with $J^{\prime} = 0.01$. In Fig.~\ref{fig:DOS_vac}(a), the TRS-breaking fields are chosen as $(\kp, \eta, \chi) = (0.08, -0.02, 0.0)$, such that in the clean model, all three bands are gapped and two of them are equivalent. In the presence of vacancies, additional modes are accumulated at low energies below the bulk gaps. Some of these vacancy-induced modes remain as delocalized as the bulk, which extends the bulk DOS to lower energies. However, when close to zero energy, the vacancy-induced modes become much more localized, as indicated by the sudden increase in IPR. In Fig.~\ref{fig:DOS_vac}(b), the parameters $(\kp, \eta, \chi) = (0.08, -0.08, 0.0)$ are chosen on a Chern number phase boundary, where one of the three bands becomes gapless (blue curve). This implies that some bulk modes could leak into the low-energy region. As we will show in the next section, this gives rise to higher instabilities on the topological property in the presence of vacancies.

\begin{figure*}
\includegraphics[width=1.0\linewidth]{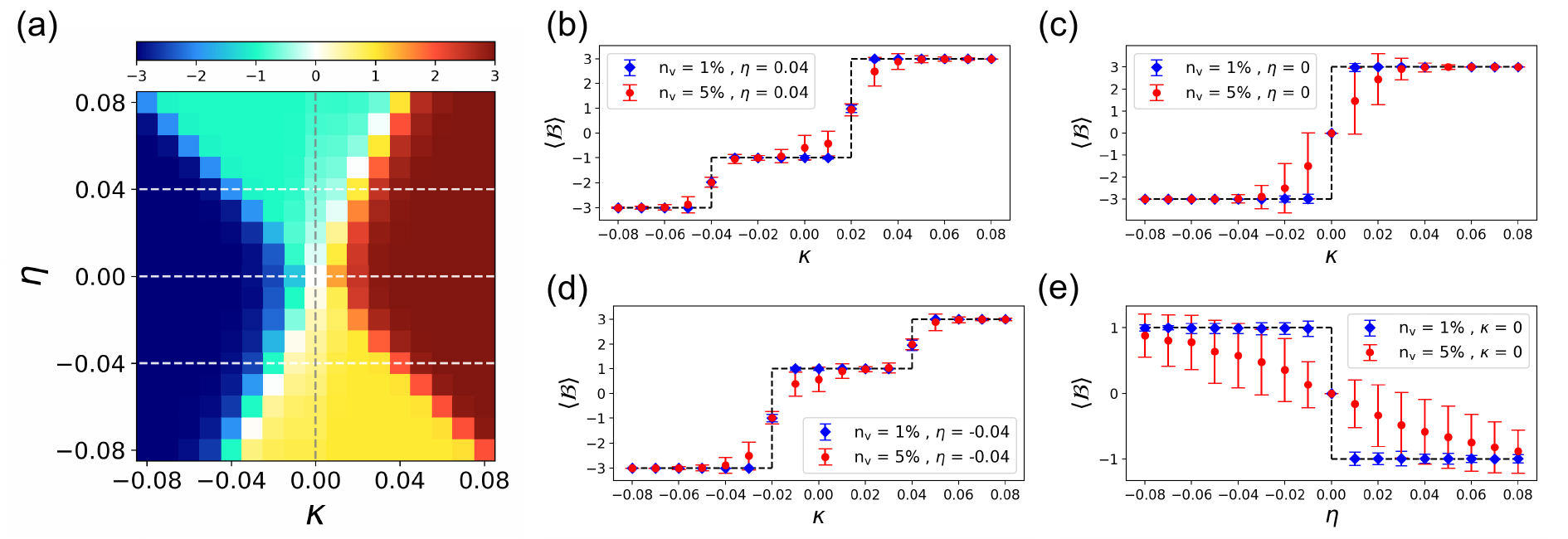}
\caption{\label{fig:Bott}  Site-disorder effects on the topological index. (a) Phase diagram of disorder-averaged Bott index of site-diluted systems in the bound-flux sector with $n_v = 5\%$ vacancy concentration. The three horizontal dashed lines and the vertical dashed line on the phase diagram indicate the four lines in (b-e). (b-e) Disorder-averaged Bott index under different vacancy concentrations in the bound-flux sector. In all the plots, each data point is calculated from 1000 random samples with $L = 20$ and $J^{\prime} = 0.01$. The error bars indicate the standard deviation of the 1000 integer-valued Bott indices. All energy scales are given in the units of $J$ and all the second-neighbor hopping around the vacancies are set to zero {(see Appendix \ref{flux binding})}. }
\end{figure*}

\subsection{Effect of vacancies on topology}

\begin{figure}
\includegraphics[width=0.9\linewidth]{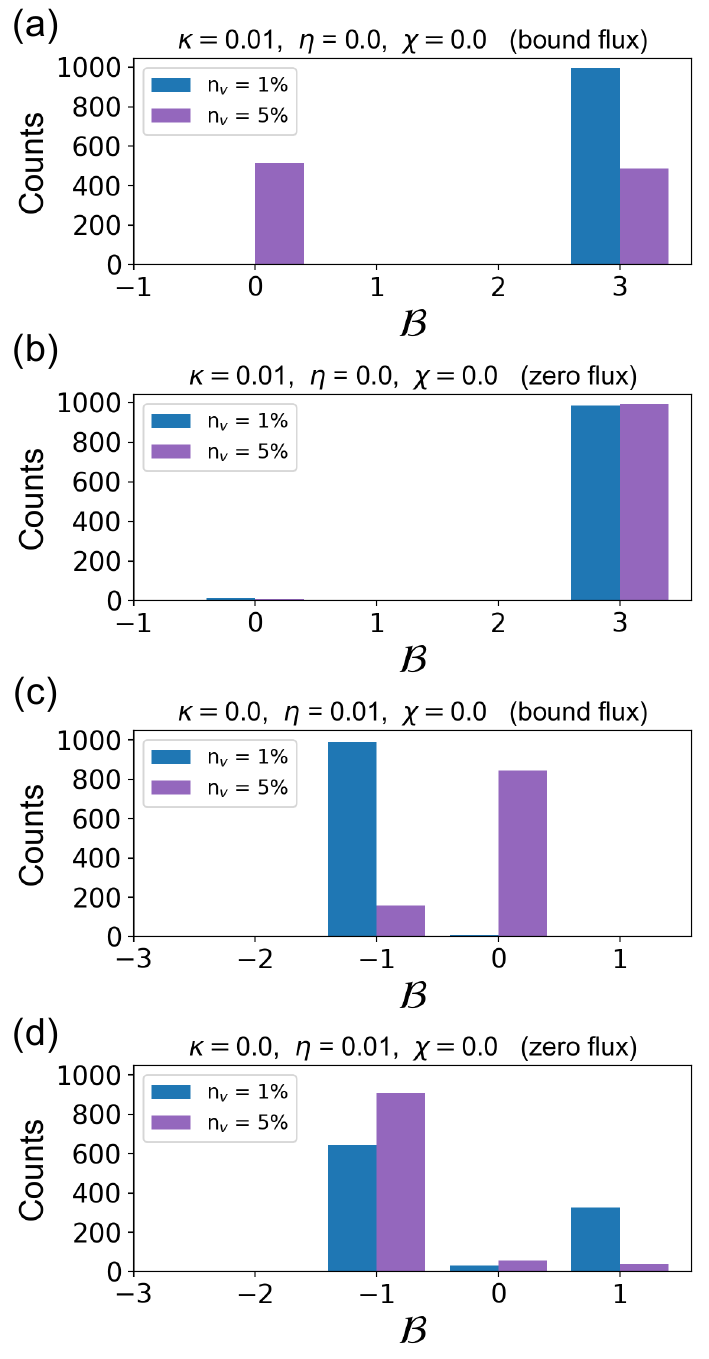}
\caption{\label{fig:Bott_hist} 
The bar chart of the Bott index for different flux sectors and vacancy concentrations. In (a-b), $\kp$ is the only second-neighbor hopping and the ideal topological index in the clean system is $+3$. In (c-d), $\eta$ is the only second-neighbor hopping and the ideal topological index in the clean system is $-1$. All the data points have integer-valued Bott index and the value associated with each bar is shown in the nearest tick of the horizontal axis. The details of the numerical calculation are the same as in Fig.~\ref{fig:Bott}.}
\end{figure}

It is known that topological phase transition occurs when the gap in the energy spectrum closes. That is why one can assume that low-energy modes from the vacancies can close the gap and consequently shift
phase boundaries. However, vacancy-induced modes are also capable of forming additional low-energy bands and increasing the band topological index. This scenario is similar to the clean Kitaev model under TRS breaking with a lattice of fluxes \cite{Lahtinen2010,Lahtinen2012,Shangshun2019,Shangshun2020}. Therefore, the disorder effect on the topological index of the Yao-Lee model can be quite perplexing due to the interplay of multi-band interactions, localized eigenmodes, and emergent low-energy bands.

{ In the presence of vacancies, the system lacks the lattice translational symmetries and thus the direct calculation of the Chern number in momentum space is impossible. In order to understand the impact of vacancies on the topology of the Yao-Lee model, we instead compute the Bott index in the real space. The Bott index measures the non-commutativity of the projected position operators and thus can be utilized in identifying topological phases~\cite{Hastings2010,Hastings2011,Loring2010, Toniolo2018,Toniolo2022}. In a system with short-range interactions and gapped extended states, the Bott index provides the integer-valued topological invariant even for small system sizes. Therefore, it is an ideal indicator of topology and has been used extensively in the models with random hoppings or potentials, quasicrystal lattices, and amorphous systems~\cite{Huang2018PRL,Huang2018PRB, Agarwala2017, Wang2020PRL, Dantas2022}. The Bott index is defined as:
\begin{eqnarray}
     \mathcal{B}=\frac{1}{2\pi}{\rm Im} {\rm Tr}
     \log (VUV^{\dagger}U^{\dagger}),
\end{eqnarray}
where $U$ and $V$ are the almost-commuting matrices defined as
\begin{align}
\begin{split}
Pe^{2\pi i R_x/L}P = T\bg 0 & 0 \\ 0 & U \ed T^{\dg}\\
Pe^{2\pi i R_y/L}P = T\bg 0 & 0 \\ 0 & V \ed T^{\dg}.\\
\end{split}
\end{align}
In the above expression, $P$ is the projection operator onto the occupied states
\begin{align}
P = T\bg 0 & 0 \\ 0 & \mathbb{1}\ed T^{\dg}
\end{align}
and the matrix $R_x$ ($R_y$) is a diagonal matrix with the entries equal to the unit-cell positions $x_i$ ($y_i$). The Bogoliubov transformation matrix $T$ is defied in Eq.(\ref{XY}).}



In the presence of vacancies, we calculate the Bott index averaged over disorder realizations. For each random sample, the Bott index is an integer but its value depends on the random configuration. In Fig.~\ref{fig:DOS_vac}, we also present the { bar chart} of the Bott-index results over 500 realizations for each vacancy concentration. When the parameters are located far away from the Chern-number phase boundary, the Bott-index results are quite robust and consistent with the Chern number of the clean model {(see Fig.~\ref{fig:phase diagram} and Fig.~\ref{fig:DOS_vac})}. On the other hand, when the parameters are close to the phase boundaries, some random samples can have Bott indices different from the ideal value of the clean model, indicating the weak instability of the Majorana band topology.
 
{ In Fig.~\ref{fig:Bott}(a) we show the topological phase diagram from the disorder-averaged Bott index of the $5\%$ site-diluted YL model with the bound-flux sector. It is apparent that the general features are consistent with the Chern-number phase diagram of the clean model (Fig.~\ref{fig:phase diagram}), and the deviation only happens near the critical lines. In Fig.~\ref{fig:Bott}(b-e), we choose four lines to demonstrate its dependence on the vacancy concentration. Generally speaking, a higher concentration of vacancies makes the disorder-averaged Bott index farther away from the ideal value of the clean model, and this effect is more discernible in the proximity of the critical points due to the diminishing of the band gaps. However, when the parameters are right at the critical values, the standard deviation of the Bott index is much smaller than  in the proximity to the critical lines. This is due to the fact that some of the bands are already closed, 
while the other bands remain gapped and stable (Fig.~\ref{fig:Bott}(b,d)). By comparing the cases with only one type of second-neighbor hopping (Fig.~\ref{fig:Bott}(c,e)), we see that the $\kp$-term leads to more robust topological phases than the $\eta$-term. This can be understood by the band energy in the clean limit (see Eq.~(\ref{spectrum})), which shows that in the case of $\eta$-term only, two bands have smaller band gaps compared to the case with only the $\kp$-term in the same magnitude. 

Even though the bound-flux sector is the ground-state flux sector in the site-diluted model, the ground-state energy difference between the bound- and zero-flux sector is quite small (see Appendix \ref{flux binding}). Therefore, it is worth discerning their impacts on the topology. In Fig.~\ref{fig:Bott_hist}, we consider this flux-sector effect in the $\kp$-term only and $\eta$-term only cases. When the $\kp$-term is the only presented TRS-breaking field, the spectrum is equivalent to three replicas of the Majorana bands in the Kitaev model, such that the Bott index in the clean limit is $+3$. At low vacancy concentration ($n_v = 1\%$), most random samples exhibit this ideal value. However, at higher vacancy concentration ($n_v = 5\%$), the nontrivial band topology can be easily shattered in the bound-flux sector, while it remains robust in the zero-flux sector (Fig.~\ref{fig:Bott_hist}(a,b)). This is due to the fact that vacancy-induced modes in the bound-flux sector can be accumulated around $E = 0$, while in the zero-flux sector there exists a small gap set by $J'$. As the vacancy concentration increases, this gap can be slightly enlarged due to the hybridization of the vacancy-induced modes, which is recently reported in the site-diluted Kitaev model\cite{Kao2024short,Kao2024long}.

On the other hand, when the $\eta$-term is the only TRS-breaking field, two bands have a smaller gap and the third band has a larger gap, according to Eq.~(\ref{spectrum}). In the bound-flux sector, $n_v = 5\%$ is enough to destroy the topology of the whole system in most of the samples (Fig.~\ref{fig:Bott_hist}(c)). However, in the zero-flux sector, the topological behavior seems more perplexing. There, low vacancy concentration of $1\%$ is enough to close the gap for the two bands with topological index $-1$, such that the resulting total topological index is $+1$. As $n_v$ increases, the topology of the three bands again becomes more robust and most of the random samples exhibit total topological index $-1$ (Fig.~\ref{fig:Bott_hist}(d)). Comparing the Bott-index results in Fig.~\ref{fig:Bott_hist}(c) and (d), the dependence on the vacancy concentration is again distinct between the bound-flux and the zero-flux sectors. This result highlights the significant role of vacancy-induced fluxes on the robustness of the band topology. }

\section{Conclusions}\label{sec:conclusion}
In this work, we investigated the Yao-Lee spin-orbital model, Eq. (\ref{eq: Hamiltonian}), in the presence of time-reversal symmetry-breaking fields and vacancies. 
We presented three types of perturbations to the original model, that split formerly three-fold degenerate spectrum into distinct bands.
For these perturbations, we found an exact phase diagram, revealing different topological regions, separated by nodal lines and characterized by the total Chern number. This phase diagram offers a comprehensive understanding of the system's topological properties.

Upon introducing vacancies to the system, the ground-state flux sector transforms into the bound-flux sector, where each lattice plaquette containing a vacancy binds a flux. {The stability of the bound-flux sector in the presence of vacancies depends on the details of the effective TRS-breaking fields around the vacancies (see Appendix \ref{flux binding})}.
Similar to the Kitaev honeycomb model, we found that in the presence of vacancies, the Majorana density of states exhibits a pileup of low-energy states, which can lead to observable effects such as an upturn in the specific heat at low temperatures \cite{Kao2021vacancy}. 
While the structure of the low-energy spectrum depends on the details of the model, such as the strength and nature of the time-reversal symmetry-breaking fields, the states in this pileup are more localized than most of the other states of the system, which is evident from the numerical simulations and can be seen from the distribution of the inverse participation ratio.

{To study the impact of the vacancies on topology, we numerically calculated the Bott index for each disordered realization, which is analogous to the Chern number but in the real-space formalism. Away from the critical lines of the topological phase diagram, the disorder-averaged Bott index is consistent with the Chern number in the clean model with an analytical solution. In close proximity to the critical lines, even a small amount of vacancies can lead to different values of the Bott index, so that the disorder-averaged value becomes non-integer. Furthermore, despite the tiny energy difference between the bound-flux and zero-flux sectors, their impact on the disorder-averaged Bott index and corresponding dependence on vacancy concentration is noticeable.}

{ Our results conclude that while a low concentration of vacancies in the extended Yao-Lee model can lead to a drastic change in the Majorana density of states, its impact on the topological property is quite limited. Nevertheless, near the critical lines, both site disorder and flux disorder play a significant role in changing the topological index of the system, and the latter effect could be more important when considering the thermal fluctuations and the TRS-breaking fields beyond the exactly-solvable limit, which requires further investigations of the extended Yao-Lee model.} 


  \vspace*{0.3cm} 
\noindent{\it  Acknowledgments:} 
We thank Alexei Tsvelik for bringing our attention to the Yao-Lee model and Vitor Dantas for the useful discussions.
Our work was supported by the  National Science Foundation  under Award No.\ DMR-1929311.   
N.B.P. also
acknowledges the hospitality and partial support of the Technical University of Munich – Institute for Advanced Study and the support of the Alexander von Humboldt Foundation.

\appendix

{\section{Perturbation theory}\label{pert}
The  Yao-Lee model (\ref{eff_ham}) has a zero-flux sector as the ground state with the flux gap $\Delta_F$. 
 In this appendix,  we show how to obtain
the  effective Hamiltonian  (\ref{eq: Hamiltonian}) by treating the  term $V$  from \eqref{pertValphabeta} perturbatively.
Consider the Hamiltonian 
\begin{equation}
    H=H_{YL}'+V,
\end{equation}
where $H_{YL}'$ and $V$ are given by equations \eqref{eff_ham} and  \eqref{pertValphabeta}, respectively. For convenience, we write $  V= V_\tau +V_\sigma$, where 
   $V_\tau=\sum_i (h_x \tau^x_i +h_y \tau^y_i +h_z \tau^z_i)$  and  $V_\sigma=K \sum_{\langle\langle ik\rangle\rangle} \boldsymbol{\sigma}_i \boldsymbol{\sigma}_k$.

 Let us compute the first four orders of the perturbation theory. They  have the following form:
\begin{align}\label{greenpert}
\begin{split}
&H^{(1)}_{\rm{eff}}=\langle 0|\Gamma V \Gamma |0\rangle\\
&H^{(2)}_{\rm{eff}}=\langle 0|\Gamma V G_0' V \Gamma |0\rangle \\
&H^{(3)}_{\rm{eff}}=\langle 0|\Gamma V G_0' V G_0' V \Gamma |0\rangle\\
&H^{(4)}_{\rm{eff}}=\langle 0|\Gamma V G_0' V G_0' V G_0' V \Gamma |0\rangle 
\end{split}
\end{align}
where $|0\rangle$ is the ground state of $H_0=H_{YL}'$ and $G_0'=\frac{1-|0\rangle \langle 0|}{E_0-H_0}$ is the resolvent. It vanishes in the zero-flux sector and acts as Green's function of $H_0$ on all other states. $\Gamma$ is the projector on the zero-flux sector.
In the following,  we rewrite $V$ using the Majorana-fermion representation for the Pauli matrices \eqref{PauliMajorana} and assume that operators quadratic in $c$  always have a more significant impact on the low-energy physics than the ones quartic in $c$, even if they originate from higher orders of perturbation theory.   \par

Next, we examine the effective Hamiltonian order by order.
The first-order term, $H_{\text{eff}}^{(1)}$, can be omitted because
  $\tau$ operators in $V_\tau$ create fluxes and are therefore annihilated by $\Gamma$, and 
  the operators $\sigma_i^\lambda \sigma_k^\mu$ in $V_\sigma$ lead to terms quartic in $c$.

 The second-order term, $H_{\rm{eff}}^{(2)}$, consists of three terms. The first one, $\tau_i^\alpha \tau_j^\beta$,  is a constant if $i=j$ and $\alpha=\beta$. If $i$ and $j$ form a bond $\langle ij \rangle_{\alpha}$, acting with operators $D_i$ and  $D_j$ (with $D_i=-ic_i^xc_i^yc_i^zd_i^xd_i^yd_i^z$, see \cite{YaoLee2011})  one can show that this term is sixth-order in $c$. In all other cases, $H_{\rm{eff}}^{(2)}$ creates fluxes and gives zero contribution. The second term  
 in $H_{\rm{eff}}^{(2)}$ given by $\sigma_i^\lambda \sigma_k^\mu \tau_k^\alpha$ can be ignored for the same reasons as the first order term. 
 The third term, $\sigma_i^\lambda \sigma_k^\mu\sigma_m^\alpha \sigma_l^\beta$, is zero because $G_0'=0$ for the zero-flux sector.\par

In computing $H_{\rm{eff}}^{(3)}$, each of the  $V$ terms  in \eqref{greenpert}  should be  chosen as either $V_\tau$ or $V_\sigma$, and  
because of $G_0'$ we cannot pick $V_\sigma$ for all three times.
The combination of one $V_\tau$ and two $V_\sigma$ terms does not preserve the flux sector, so it gives zero. The combination of two $V_\tau$ and one $V_\sigma$ does not create fluxes if $\tau$ operators are on adjacent sites with link of  $\alpha$-type but, after applying $D$ operator twice, we get $\tau_i^\alpha \tau_j^\beta=i u_{ij}c_i^x c_i^y c_i^z c_j^x c_j^y c_j^z$. The remaining operator $\sigma_l^\lambda \sigma_k^\mu$ in the best case will eliminate two $c$ operators on one site 
$l$ (if $l=i$ or $l=j$) but will add two more $c$ operators on site $k$. Thus, this is at least the sixth order in $c$. The last possible case is that all the terms  are formed by  $V_\tau$ operators. In order to preserve the zero-flux sector one needs to pick a site and three bonds connected to it and set the term in $H_{\rm{eff}}^{(3)}$ to have a form $\tau_i^x\tau_j^y\tau_k^z$ where each index 
$i$, $j$, $k$ is chosen among two sites of bonds $x,y,z$, respectively. It is easy to see that the term $\tau_i^x\tau_i^y\tau_i^z$ gives a trivial contribution, and all the other possibilities are at least sixth order in $c$. \par

For the same reasons as for $H_{\rm{eff}}^{(3)}$, computing $H_{\rm{eff}}^{(4)}$, we cannot set all $V$ terms to be $V_\sigma$. By iterating all possible combinations one can also show that the perturbation with two $V_\tau$ and two $V_\sigma$ operators gives at least sixth order in the $c$ operators. The same is true for all four $V$ chosen to be $V_\tau$. 
Finally, the perturbation with one $V_\sigma$ and three $V_\tau$ can give contribution quadratic in $c$ if it has the form $\tau_i^x \tau_j^z \tau_k^y (\sigma_i^\alpha \sigma_k^\beta)$, where $ij$ is $x$-bond and $jk$ is $y$-bond, or any other configurations equivalent to this one. Rewriting this expression in terms of Majorana fermions gives
\begin{align}\label{secondneighbor}
    \tau_i^x \tau_j^z \tau_k^y (\sigma_i^\alpha \sigma_k^\beta)=i u_{ij}u_{jk}c_i^\alpha c_k^\beta .
\end{align}
\par

Now let us define the prefactor for this term.  Acting by $V_\sigma$  does not create fluxes, and due to the resolvent $G_0'$, it cannot occupy the first or the last place in \eqref{greenpert}. The remaining possibilities give
\begin{widetext}
\begin{align}\label{h4}
    H^{(4)}_{\rm{eff}}=\sum_{ijmnl}\sum_{\mu \nu \alpha \beta \gamma}(h_i^\alpha \tau_i^\alpha)\frac{1}{\Delta_{jm}^{\beta \gamma}}(h_j^\beta \tau_j^\beta)\frac{1}{\Delta_{m}^\gamma} K^{\mu \nu}\sigma_n^\mu\sigma_l^\nu\frac{1}{\Delta_m^\gamma}(h_m^\gamma \tau_m^\gamma)+ 
    (h_i^\alpha \tau_i^\alpha)\frac{1}{\Delta_{jm}^{\beta \gamma}}K^{\mu \nu}\sigma_n^\mu\sigma_l^\nu\frac{1}{\Delta_{jm}^{\beta \gamma}}(h_j^\beta \tau_j^\beta) \frac{1}{\Delta_m^\gamma}(h_m^\gamma \tau_m^\gamma),
\end{align}
\end{widetext}
where $\Delta_m^\gamma$ is the change of energy after applying $\tau_m^\gamma$ operator and $\Delta_{jm}^{\beta \gamma}$ is the same quantity but after applying the $\tau_j^\beta \tau_m^\gamma$ operator. In the absence of vacancies, all $\Delta$ are simply the energies of creating two adjacent fluxes.
Expression of the form \eqref{secondneighbor} can be obtained from each of the two terms of \eqref{h4} by choosing from which term $\tau^\alpha$ will come. Since $\tau$ operators on different sites commute with each other, all possible permutations of them give the same result. For that reason there are $2 \cdot 3!$ possible contributions, which gives \eqref{diagonal perturbation}.
    
}

 \begin{figure*}
     \includegraphics[width=1\linewidth]{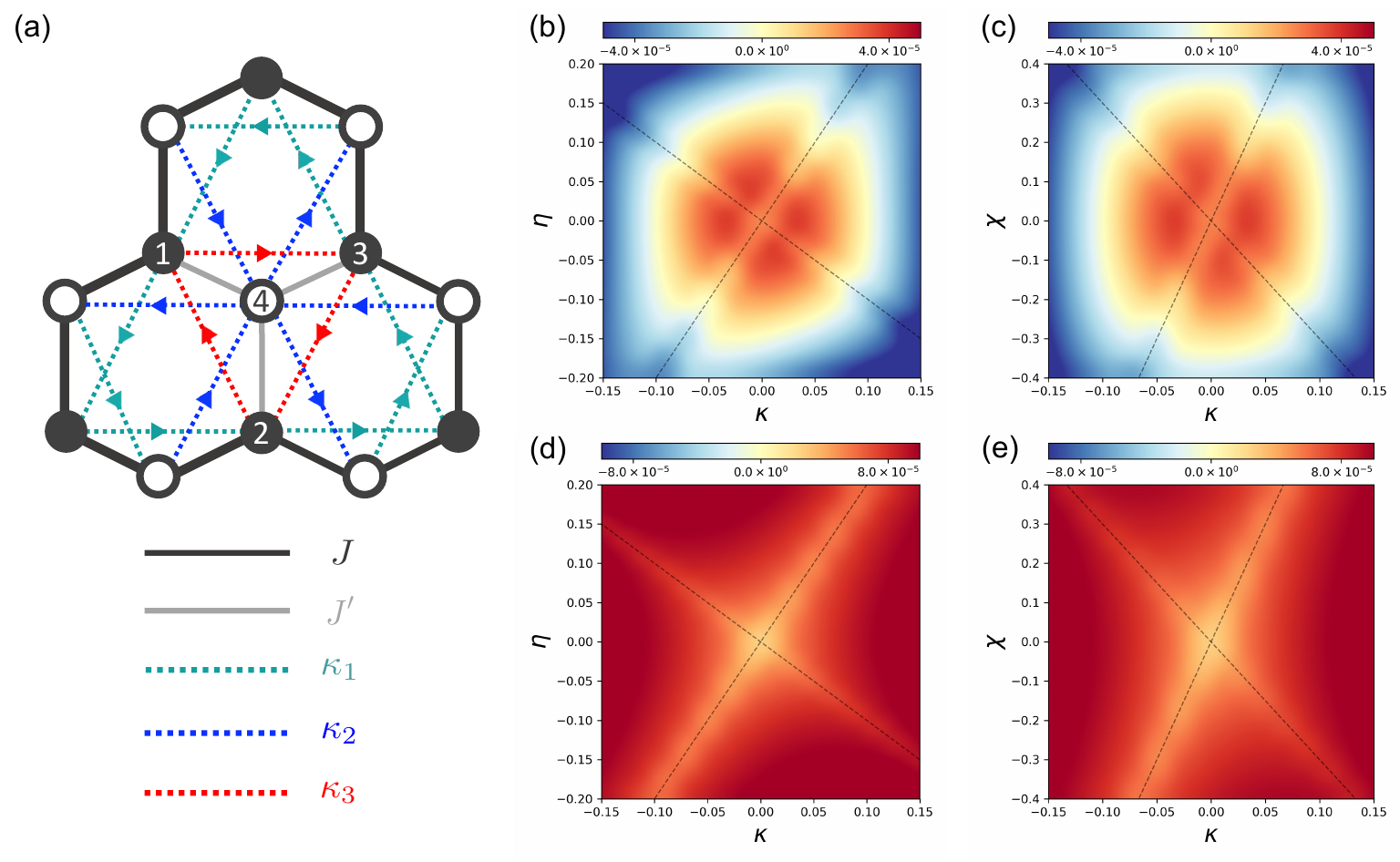}\caption{\label{fig:BE} { (a) Second-neighbor hoppings around the vacancy site (site 4). In the presence of weak couplings $J'$ (gray lines), the second-neighbor hopping strength $\kappa$ is divided into three types: $\kappa_1$ (green dotted lines), $\kappa_2$ (blue dotted lines), and $\kappa_3$ (red dotted lines). The couplings $\eta$ and $\chi$ are also classified into three types. (b-c) The phase diagrams of $E_{\mathrm{diff}}$ as a function of $\kappa$, $\eta$, and $\chi$. We set $\kappa = \kappa_1 = \kappa_2 = \kappa_3$, $\eta = \eta_1 = \eta_2 = \eta_3$, and $\chi = \chi_1 = \chi_2 = \chi_3$. (d-e) The phase diagrams with $\kappa = \kappa_1$, $\eta = \eta_1$, and $\chi = \chi_1$. All the other components are set to zero. The phase diagrams are calculated from the $L=40$ system with periodic boundary conditions and $J' = 0.01$. The gray dotted lines denote the critical lines of the Chern-number phase diagram. All energy scales are given in the units of $J$.}}
\end{figure*}

\section{TRS-breaking field for vacancy} \label{flux binding}
{  In the presence of a true vacancy ($J'=0$), three adjacent plaquettes merge into one and hence flipping the link variable connected to a vacancy does not change flux configuration nor the energy of the system. In this case, Eq.~\eqref{h4} yields infinity for the second-neighbor hopping amplitude emanating from the vacancy site and for the second-neighbor hopping among the nearest neighbors of the vacancy site. This implies that the perturbation theory breaks down near the vacancies. On the other hand, if $J'$ is nonzero but smaller than $J$, those effective coupling strengths emanating from and coming across the vacancy site are distinct from the normal second-neighbor couplings in the bulk. The geometry of different types of second-neighbor couplings are illustrated in Fig.~\ref{fig:BE}(a).}

When considering vacancies in the Kitaev model, it was found that vacancies bind the fluxes of the emergent $Z_2$ gauge field, resulting in the formation of the so-called bound-flux ground state~\cite{Willans2010,Willans2011,Kao2021vacancy}. As the first step, we check whether this phenomenon persists for vacancies in the Yao-Lee model. 
{  To this end, we consider a finite system of $L \times L$ unit cells with two vacancies, one on the A sublattice site and the other on the B sublattice site (see Fig.~\ref{YLlattice}), which are separated by a distance $\sim L/2$, and then calculate the following quantity:
\begin{equation}\label{Eq:binding_energy}
    E_{\mathrm{diff}} = \frac{E_{\mathrm{zero}}-E_{\mathrm{bound}}}{2(2L^2)} .
\end{equation}
This quantity shows the energy difference between the zero-flux sector and the bound-flux sector per flux and per site. In the phase diagram of $E_{\mathrm{diff}}$, we consider two limiting cases for the effective couplings. In Fig.~\ref{fig:BE}(b-c), we set $\kp_1 = \kp_2 = \kp_3 \equiv \kp$, $\eta_1 = \eta_2 = \eta_3 \equiv \eta$, and $\chi_1 = \chi_2 = \chi_3 \equiv \chi$. In this scenario, the ground-state flux sector tends to be the bound-flux sector when the time-reversal symmetry-breaking fields are small, while the zero-flux sector becomes energetically favored once these fields surpass some critical values. Furthermore, $E_{\mathrm{diff}}$ decreases in the proximity of the critical lines (dashed lines in Fig.~\ref{fig:BE} (b-c)) because some of the Majorana bands become gapless.

Despite the decrease in $E_{\mathrm{diff}}$, the ground state remains in the bound-flux state ($E_{\mathrm{diff}}>0$) along these dashed lines before it goes to the zero-flux region with larger TRS-breaking field strengths. This scenario exhibits one clear flux-sector transition when monotonically increasing any of the three TRS-breaking fields.
 
In Fig.~\ref{fig:BE}(d-e), we consider the second scenario where only the normal effective couplings ($\kp_1$, $\eta_1$, and $\chi_1$) are present. It is the more realistic case for small $J'$ because the flux gaps associated with these weak bonds are also small and thus the perturbation theory breaks down. In this case, the ground state remains in the bound-flux sector and no flux-sector transition is observed. This is the scenario used in Sec.~\ref{sec:vacancy}.}

\bibliography{Ref.bib}

\end{document}